\documentclass[structabstract]{aa}
\usepackage{graphicx}
\usepackage{url}
\usepackage{natbib}
\usepackage{listings}
\usepackage{verbatim,amssymb}
\usepackage{longtable}

\title{Probing the cosmic-ray content of galaxy clusters by stacking Fermi-LAT count maps}
\authorrunning{B. Huber et al.}
\titlerunning{Probing the cosmic-ray content of galaxy clusters by stacking Fermi-LAT count maps}

\author{B.~Huber\inst{\ref{inst1},\ref{inst2}}
\and C.~Tchernin\inst{\ref{inst2}} 
\and D.~Eckert\inst{\ref{inst2}}
\and C.~Farnier\inst{\ref{inst2},\ref{inst3}}
\and A.~Manalaysay\inst{\ref{inst1}}
\and U.~Straumann\inst{\ref{inst1}}
\and R.~Walter\inst{\ref{inst2}}}

\institute{Physik - Institut, Universit\"at Z\"urich, Winterthurerstrasse 190, CH-8057 Z\"urich, Switzerland\label{inst1}\\
\email{ben.huber@physik.uzh.ch}
\and Department of Astronomy, University of Geneva, ch. d'Ecogia 16, CH-1290 Versoix, Switzerland\label{inst2}\\
\email{celine.tchernin@unige.ch}\\
\email{dominique.eckert@unige.ch}
\and The Oskar Klein Centre for Cosmoparticle Physics, Department of Physics, Stockholm University, Albanova,\\ SE-10691 Stockholm, Sweden\label{inst3}\\
\email{christian.farnier@fysik.su.se}}

\date{In preparation, April 2013}

\abstract {} 
{Observation in radio have shown that galaxy clusters are giant reservoirs of cosmic rays (CR). Although a gamma-ray signal from the cluster volume is expected to arise through interactions of CR protons with the ambient plasma, a confirming observation is still missing.} 
{We search for a cumulative gamma-ray emission in direction of galaxy clusters by analysing a collection of stacked Fermi-LAT count maps.
Additionally, we investigate possible systematic differences in the emission between cool-core and non-cool core cluster populations.} 
{Making use of a sample of 53 clusters selected from the HIFLUGCS catalog, we do not detect a significant signal from the stacked sample. The upper limit on the average flux per cluster derived for the total stacked sample is at the level of a few $10^{-11}\,\mathrm{ph}\,\mathrm{cm}^{-2}\,\mathrm{s}^{-1}$ at 95\% confidence level in the 1-300 GeV band, assuming power-law spectra with photon indices 2.0, 2.4, 2.8 and 3.2. Separate stacking of the cool core and non-cool core clusters in the sample lead to similar values of around $5\times 10^{-11}\,\mathrm{ph}\,\mathrm{cm}^{-2}\,\mathrm{s}^{-1}$ and $2\times 10^{-11}\,\mathrm{ph}\,\mathrm{cm}^{-2}\,\mathrm{s}^{-1}$, respectively.} 
{Under the assumption that decaying $\pi^0$, produced in collisions between CRs and the ambient thermal gas, are responsible for the gamma-ray emission, we set upper limits on the average CR content in galaxy clusters. For the entire cluster population, our upper limit on the gamma-ray flux translates into an upper limit on the average CR-to-thermal energy ratio of 4.6\% for a photon index of 2.4, although it is possible for individual systems to exceed this limit. Our 95\% upper limits are at the level expected from numerical simulations, which likely suggests that the injection of CR at cosmological shocks is less efficient than previously assumed.}


\begin{document}
\maketitle
%\linenumbers

%Begin the section.
\section{Introduction}
\label{introduction}
Galaxy clusters are the largest gravitationally bound structures in the Universe and have been successfully observed in radio, optical, UV and X-ray wavelengths \citep[see e.g.][for reviews]{sarazin86,voit05}. These observations indicate that galaxy clusters are not only large-scale accumulations of galaxies, gas and dark matter but also giant reservoirs of relativistic cosmic rays (CRs), i.e. relativistic electrons and protons, confined in the cluster volume through large-scale magnetic fields \citep{voelk96,berezinsky97,voelk99}. CRs are accelerated in shock waves induced by cluster mergers and during the accretion of material from the cluster environment. These shocks also contribute to the thermalization of substructures in the hot intra-cluster gas. CRs can also be injected into the cluster volume by central active galactic nuclei (AGN) and supernovae (SNe). 

The presence of relativistic electrons is demonstrated by observations in the radio band due to large-scale synchrotron radiation \citep[e.g.][]{feretti05,ferrari08,brunetti11,feretti12} and possibly by observed emissions in the extreme UV and hard X-ray range attributed to inverse-Compton scattering (ICS) with photons of the cosmic microwave background (CMB) \citep[e.g.][]{rephaeli08,eckert08,nevalainen09,ajello10}. CR protons are expected to be confined in galaxy clusters for very long timescales, which may lead to proton-proton ($p$-$p$) collisions between CR protons and the ambient thermal plasma. Gamma rays may then arise from the decay of neutral pions produced in these interactions \citep[e.g][]{blasi07,pfrommer08,pinzke10}, with an interaction rate for $p-p$ collisions that is expected to be highest close to the gravitational center of the cluster where the target proton density is enhanced.

In addition, accretion shocks, expected to be powerful enough to boost electrons up to 100~TeV and protons up to 1000~TeV, could be an alternative source of high-energy gamma rays \citep{vannoni09,timokhin04}. At these energies, primary and secondary electrons could undergo ICS and transform CMB photons into high-energy gamma rays \citep[e.g.,][]{blasi07}. A further possibility considers ultra-high-energy protons with energies $>10^6$ TeV, that interact with the CMB producing electron-positron pairs leading to hard X-ray and gamma-ray emission via synchrotron and ICS \citep{inoue05}.

Galaxy clusters are also strong candidates to search for an exotic signature, since a large quantity of dark matter is attested in galaxy clusters. Several observational evidences suggest that dark matter could be formed of weakly interactive massive particles. A potential gamma-ray signal observable with the Fermi-LAT could then arise from the annihilation of dark-matter particles \citep[e.g.,][]{Cola06,jeltema09,pinzke11}. 

Numerous observational studies \citep[e.g.][]{reimer03,perkins08,hess09,magic10,fermi10,COMA,Dutson12} have resulted in upper limits on the gamma-ray emission from clusters of galaxies, but a definite observation is still missing \citep[see also][]{pinzke11}.

In this work, we search for gamma-ray emission from galaxy clusters using the stacking method described in \citet{huber12}, applying a maximum likelihood analysis on stacked count maps of galaxy clusters obtained with the Large Area Telescope (LAT). 
The LAT is the main instrument aboard the Fermi Gamma Ray Space Telescope spacecraft. It is a pair conversion telescope with an effective area of $\sim 1~\mathrm{m}^{2}$ and a field of view of about $2.4~\mathrm{sr}$, generally operating in survey mode and providing an all-sky coverage every two orbits. The instrument is sensitive to gamma-ray events with energies between 20~MeV to 300~GeV \citep{atwood09}.

We take into account that galaxy clusters can be subdivided into two classes, cool-core (CC) and non-cool core (NCC) galaxy clusters \citep{cavagnolo09}, which may correspond to different stages of a cyclical cluster evolution \citep[e.g.][]{rossetti11}. In this evolutionary scenario, CC galaxy clusters are relaxed systems that host active galactic nuclei (AGN) at their center, powered by the accretion of intra-cluster gas \citep[e.g.,][]{mcnamara07,mcnamara12,fabian12}. NCC galaxy clusters, on the other hand, are disturbed systems in which merger events lead to particle acceleration \citep[e.g.][]{brunetti09,cassano10,rossetti11} through large-scale shocks \citep{markevitch07,markevitch10}.

The paper is structured as follows. In Sect.~\ref{selection}, we describe the selection of galaxy clusters that we use to search for gamma-ray emission. The data processing and analysis method used to stack the emission from the selected galaxy clusters is described in Sect.~\ref{method}. In Sect.~\ref{results}, we present the results on the gamma-ray emissivity obtained for the entire cluster sample and separately for the CC and NCC subsamples. In Sect.~\ref{sec:CRenergy}, we use our observational results to set constraints on the average CR energy density in galaxy clusters. We conclude the paper with a discussion in Sect.~\ref{sec:discussion}. 

\section{Cluster selection}
\label{selection}
From the extensive catalog of high X-ray luminosity galaxy clusters HIFLUGCS \citep{reiprich02}, a list of 53 clusters (Table~\ref{clustertable}) has been retained to fulfill selection criteria that would lead to an enhanced signal-to-noise ratio. Our selection criteria are:
\begin{itemize}
\item{} Due to the decrease of gamma-ray flux with increasing distance of the emitting object, only galaxy clusters located at low redshift $z<0.2$ have been retained.
\item{} To avoid false signal detection due to a possible mis-modeling of the observed strong diffuse emission along the galactic plane and in the galactic centre region, galaxy clusters located at latitudes $|b|<25^\circ$ and longitudes $-30<l<+30^{\circ}$ have been discarded. Additionally, due to the strong emission detected around the Taurus molecular cloud, the galaxy cluster A400 was also excluded.
\item{} Finally, to avoid false signals from residual emission of powerful gamma-ray sources listed in the $2^{\mathrm{nd}}$ year Fermi-LAT (2FGL) catalog \citep{nolan12} in the close vicinity of the galaxy clusters, A1650, A1651, A1689, A2065, A2199, A2589, A3376, HCG94, M49, NGC4636, UGC03957, and ZwCl1215 have also been removed from the selection.
\end{itemize}

\section{Analysis}
\label{method}

\subsection{Data preparation}

In this study, we make use of the gamma-ray events collected by the Fermi-LAT satellite from 2008-08-04 to 2013-01-31 analysed using the version \texttt{v9r27p1} of the Fermi Science Tools\footnote{\url{http://fermi.gsfc.nasa.gov/ssc/data/analysis/scitools/}} in conjunction with the \texttt{P7SOURCE\_V6} instrument response functions. To reduce the large number of events due to diffuse emissions and improve the point spread function of the instrument, an energy threshold of 200~MeV was applied to the event selection.
For each source listed in Table~\ref{clustertable}, \texttt{SOURCE} class photon-like events were extracted from a circular region of interest (ROI) with $\sim 10^{\circ}$ radius centered on the galaxy cluster.
Good time intervals were generated using the recommended selection expression\footnote{\texttt{(DATA\_QUAL==1) \&\& (LAT\_CONFIG==1) \&\& ABS(ROCK\_ANGLE)<52}.} and ROI-based maximum zenith angle cut was applied to exclude photons coming from the Earth limb.
A binned likelihood analysis is first performed on each individual ROI to determine the parameters of the gamma-ray emission model. The expected gamma-ray signal within the ROI is modelled using a combination of the galactic\footnote{\url{http://fermi.gsfc.nasa.gov/ssc/data/analysis/software/aux/gal_2yearp7v6_v0.fits}} (GB) and isotropic\footnote{\url{http://fermi.gsfc.nasa.gov/ssc/data/analysis/software/aux/iso_p7v6source.txt}} (EGB) diffuse emission models, and also incorporates all point-like sources listed in the 2FGL catalog within $20^{\circ}$ around the galaxy cluster. To prevent genuine variability or statistical fluctuations of the signal from nearby sources from affecting the analysis, the overall flux normalisations of sources within $10^{\circ}$ radius of the target were treated as free parameters during the likelihood analysis. The normalisations of the GB and EGB components were also free to vary during the fitting procedure to improve any local mis-modelling of the diffuse emission in this particular ROI. All other parameters were fixed to the best fitting values published in the 2FGL.
The parameters maximizing the likelihood function are then used to simulate all Fermi-LAT detected sources, with the exceptions of the GB and EGB components, and subtract them from the data to simplify the model used for the stacking analysis \citep[see][]{huber12}.

\subsection{Stacking analysis}
Following the procedure described in \citet{huber12}, the stacking of the sources is performed by stepwise-adding the point-source subtracted count maps of the selected ROIs. 
At each stacking step, the co-added map is analysed using a binned maximum likelihood approach \citep{mattox96}, in which two alternative model hypotheses are compared by maximising their respective likelihoods with respect to the obtained count map. The null hypothesis corresponds to gamma-ray events originating from the GB and EGB components only, whereas the second, alternative hypothesis includes an additional so-called \emph{test source}, corresponding to a point-like emission with power-law spectral shape parametrised as $$ dN/dE = N_0 (E/E_0)^{-\Gamma_{ph}}.$$
As the signal is expected to be faint, only the flux normalisation parameter is allowed to vary during the fitting procedure, whereas the photon spectral index, $\Gamma_{ph}$ was fixed to a set of values $2.0, 2.4, 2.8, \,\mathrm{and}\, 3.2$ and the energy scale, $E_0$ was set to 1~GeV. We performed the spectral fit in the energy range between 1 and 300~GeV where a power law is a good approximation of the gamma-ray spectrum resulting from neutral pion decay arising in proton-proton collisions (see Sect.~\ref{sec:CRenergy} and Appendix \ref{app:calculation}). This particular choice of gamma-ray spectral indices allows us to bracket the range of indices of the injected proton spectrum \citep{pinzke10,Vazza12a}.

The probability of the presence of the additional \emph{test source} is obtained from a likelihood ratio test defined as $$ \mathrm{TS} = -2\,\mathrm{ln}\, \frac{\mathcal{L}^{max}_{0}}{\mathcal{L}^{max}_{1}},$$ where $\mathcal{L}^{max}_{0}$ and $\mathcal{L}^{max}_{1}$ are the maximum likelihood values for the null and alternative hypotheses respectively. If the null hypothesis is true, then the detection significance of the additional source is approximately given by $\sqrt{\mathrm{TS}}$. In the following, we use a detection threshold of $\mathrm{TS}=25$ to report the flux of the source. In case of a lower $\mathrm{TS}$ value, a 95\% confidence level (CL) upper limit on the gamma-ray flux divided by the sample size (SUL/N) is reported. Namely, the corresponding upper limit represents the maximum allowed average flux per system in the sample (at 95\% CL). We stress that, since this is an average value, this limit does not hold on an individual-object basis, and it is possible for some systems to exhibit a higher gamma-ray flux. For the upper limits on individual systems, we refer the reader to \citet{fermi10}.

\subsection{Stacking order}
In hierarchical formation scenarios, galaxy clusters with large masses imply long formation history, which in turn allows CRs to be accumulated in the cluster volume over time \citep{berezinsky97}. It is therefore reasonable to expect that the gamma-ray luminosity correlates with the galaxy cluster mass \citep[see][]{pinzke10,zandanel12}.
On the other hand, since the measured flux is inversely proportional to the square of the distance of the source, similar mass objects located at large distance will be less likely detected.
For these reasons, we stack the galaxy clusters in a sorted sequence defined by descending values of $M_{500} / z^2$, where $M_{500}$ is the galaxy cluster mass encompassed in the volume for which the density is 500 times larger that the critical density of the Universe and $z$ is the cluster redshift.

We also tested other stacking orders, and found that the TS or flux upper-limits development were not importantly modified.

\section{Results}
\label{results}

In this section we present the results on the TS and flux upper limits (SUL/N) obtained for the additional point-like emission after each stacking step on the whole set of galaxy clusters listed in Table~\ref{clustertable}. Since the evolution histories of CC and NCC galaxy clusters are different we also provide results for these two populations of galaxy clusters.

\subsection{Stacking of 53 clusters}
\label{53clusters}

\begin{figure}
\resizebox{\hsize}{!}{\includegraphics[width=\linewidth]{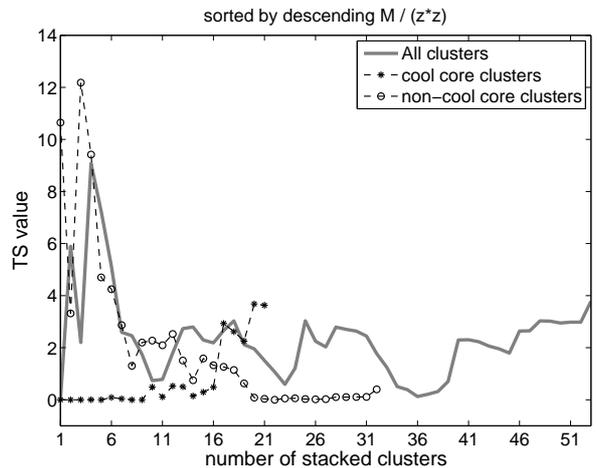}}
\caption{Evolution of the TS corresponding to the additional presence of a point-like emission at the center of the ROIs as function of the number of stacked galaxy clusters obtained from the whole sample (plain line), as well as for cool core (star-dashed line) and non-cool core (circle-dashed line) populations. The TS values are obtained for a power-law spectrum with photon index $\Gamma_{ph}=2.4$ and galaxy cluster ordered by descending values of $M_{500}/z^2$.}
\label{TS_allClusters}
\end{figure}

The TS development obtained after each stacking step of our 53 galaxy clusters order by descending $M_{500}/z^2$-values is shown in Fig.~\ref{TS_allClusters}. For every stacking step the resulting TS value remains below the detection threshold $\mathrm{TS} = 25$, with a final TS value of 3.8 after 52 stacking steps for a photon index of 2.4, not leading to any significant signal after any stacking step.

\begin{figure}
\resizebox{\hsize}{!}{\includegraphics[width=\linewidth]{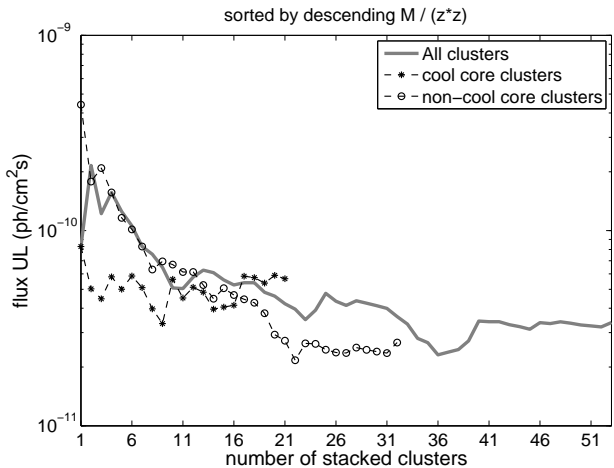}}
\caption{95\% CL upper-limits on the average photon flux (SUL/N) in the $1-300$ GeV energy band for a potential point-like emission at the center of the ROIs versus the number of stacked clusters. The stacking analysis is performed using a power-law spectrum with photon index $2.4$ and a stacking order of descending $M_{500} / z^2$-values.}
\label{FluxUL_differentIndices_allClusters}
\end{figure}

Since no significant signal was detected, we computed the corresponding 95\% CL gamma-ray flux upper limit (SUL/N) integrated between 1 and 300~GeV, shown in Fig.~\ref{FluxUL_differentIndices_allClusters}. Stacking all 53 clusters from table \ref{clustertable} and analysing them with photon indices $\Gamma_{ph}$ = 2.0, 2.4, 2.8, and 3.2 yield final upper limits on the gamma-ray emissivity of the galaxy cluster sample of the order of few $10^{-11}\,\mathrm{ph}\,\mathrm{cm}^{-2}\,\mathrm{s}^{-1}$ in the 1 to 300~GeV range. In comparison, the individual flux (SUL/N) obtained for a power-law with spectral index $\Gamma_{ph}=2.0$ and translated in the same energy range, published in \citet{fermi10} range between $1.5 \,\mathrm{and}\, 28.2\times 10^{-10}\,\mathrm{ph}\,\mathrm{cm}^{-2}\,\mathrm{s}^{-1}$, not considering the comparatively high values for the Ophiuchus and Perseus clusters.

\subsection{Separate stacking of cool core and non-cool core clusters}

In the following, we separate the CC and NCC populations from our sample of 53 clusters and perform independent analyses on the two subsamples. Due to the presence of a central radio-loud AGN in CCs \citep{mittal09}, gamma-rays could originate, for instance, from CRs ejected by the AGN, that interact with the ambient cluster gas \citep[e.g.,][]{Mathews,WR,Fujita}. 
NCCs, on the other hand, are expected to be disturbed systems with ongoing dynamical activity, that could lead to the presence of CRs accelerated through merger shocks and large-scale turbulence. 
The separate treatment of the two types of galaxy clusters could resolve intrinsic differences in their gamma-ray emissivity. The class of each cluster is indicated in Table~\ref{clustertable}, following the classification of \citet{cavagnolo09}, when available, and \citet{chen07} otherwise. Though not significant, the largest discrepancy in the resulting TS development between both samples is found for a stacking sequence defined by descending values of $M_{500} / z^2$.

\subsubsection{Stacking cool core clusters}

From our selection of galaxy clusters, 21 objects belong to the CC class. The development of the TS of the stacked CC galaxy clusters sub-sample ordered by descending $M_{500} / z^2$ is presented in Fig.~\ref{TS_allClusters}. The TS values remain below 4 during the whole stacking process and thus no significant signal was detected. The 21 stacked CC galaxy clusters result in final flux SUL/N $5.7\times 10^{-11}\,\mathrm{ph}\,\mathrm{cm}^{-2}\,\mathrm{s}^{-1}$ for an intermediate photon index $\Gamma_{ph} = 2.4$.

\subsubsection{Stacking non-cool core clusters}

We repeated the same analysis for our sample of 32 NCC galaxy clusters. The sample is again sorted by descending $M_{500} / z^2$ values, and the corresponding TS development during the stacking procedure is shown in Fig.~\ref{TS_allClusters}. In contrast to the result obtained for the CC sample, the largest value of the TS, though not statistically significant, is obtained after the stacking of few large and nearby objects. The final TS value is however below 1. The 95\% CL flux SUL/N obtained for the final sample of 32 stacked NCC galaxy clusters is $2.7\times 10^{-11}\,\mathrm{ph}\,\mathrm{cm}^{-2}\,\mathrm{s}^{-1}$ for $\Gamma_{ph} = 2.4$.
\newline

In all investigated cases, the TS values remain below the detection threshold. The final 95\% CL flux upper limits obtained for power laws with photon indices $\Gamma_{ph} = 2.0, 2.4, 2.8, \,\mathrm{and}\, 3.2$ integrated from 1 to 300~GeV, as well as the final TS values, are summarized in Table~\ref{FluxULTable}.

\begin{table*}
  \caption{\label{FluxULTable}Flux upper limits divided by the sample size (SUL/N) on the stacked emission in the $1-300$ GeV energy band. }
\small
\begin{center}
%\begin{tabular}{ccp{1.5cm}p{1.5cm}p{1.5cm}p{1.5cm}p{1.5cm}p{1.5cm}}
\begin{tabular}{ccccccccc}
\hline
Photon index  & \multicolumn{2}{c}{All} & \multicolumn{2}{c}{CC} & \multicolumn{2}{c}{NCC} \\
$\Gamma_{ph}$ & Flux UL & Final TS & Flux UL & Final TS & Flux UL & Final TS\\
\hline
\hline
2.0  & 3.2 & 8.3 & 4.9 & 5.7 & 2.8 & 2.1\\
2.4   & 3.4 & 3.8 & 5.7 & 3.6 & 2.7 & 0.4\\
2.8   & 2.9 & 1.3 & 5.7 & 2.3 & 2.0 & 0.0\\
3.2   & 2.4 & 0.3 & 5.5 & 1.8 & 1.5 & 0.0\\
\hline

\end{tabular}
\end{center}
\textbf{Column description:} 1: Photon index assumed for the analysis. 2-3: 95\% CL flux upper limit on the stacked emission of the entire sample in units of $10^{-11}$ ph cm$^{-2}$ s$^{-1}$ and final TS value after 53 stacking steps. 4-5: Same as 2-3 for the CC subsample. 6-7: Same as 2-3 for the NCC subsample.
\end{table*}

%\vspace{-0.4cm}

\section{Constraints on the cosmic-ray energy density}
\label{sec:CRenergy}
%%%%%%%%%%%%%%%%%%%%%%%%%%%%%%%%%%%%%%%%%%%%%%%%%%%%%%
The gamma-ray data explored in this paper can be combined with X-ray observations to extract constraints and limits on the averaged CR energy density in galaxy clusters. 
 Indeed, theoretical models generally agree in reporting that the gamma-ray emission from clusters should be dominated by the decay of $\pi^0$ produced in $p-p$ interactions between the CR and the ambient intracluster medium (ICM) \citep[e.g.,][]{Cola98,Pfrommer04,pfrommer08},
\begin{equation}p+p\rightarrow \pi^0 \rightarrow 2\gamma.\end{equation}
 
Assuming that the gamma-ray flux is resulting from this process, constraints on the energy density stored in CRs can be obtained. The distribution of the ambient gas for the systems is known from X-ray observations \citep[e.g.,][]{reiprich02,chen07}, and thus upper limits on the CR-to-thermal energy ratio can be derived. 

\subsection{Method}

We describe here the method used to determine upper limits on the average CR-to-thermal energy ratio, $\epsilon$, in our galaxy cluster sample. For more details about the calculation, we refer the reader to Appendix \ref{app:calculation}.

We consider a CR population with a power-law energy spectrum with index $\Gamma_p$. To cover the range of spectral indices predicted by simulations, we vary the spectral index of the underlying proton population between $\sim2$ (corresponding to the strong shock limit, i.e. Mach numbers $M\geq10$) and $\sim3.2$ (corresponding to weak shocks with $M\sim2$). Through $\pi^0$ decay, this CR population produces a gamma-ray signal with a slightly harder effective photon spectrum (see Appendix \ref{app:calculation}). We consider only the protons with an energy larger than the threshold for $\pi^0$ production. In other terms, we set a low-energy cut to the proton population at a kinetic energy of 1 GeV, since protons with lower energies are not observable. We assume that the density of the thermal gas follows an isothermal beta model \citep{Cavaliere}, with parameters for each cluster obtained from the literature \citep{chen07}. The produced $\gamma$-ray luminosity is then obtained by integrating the photon production rate over the cluster volume,

\begin{equation}
L=2\int^{R_{vir}}_{0}{n_{gas}(r)n_{CR}(r)v_{CR}\sigma_{pp} E_{\gamma}}\,\mbox{d}^3r,
\end{equation}

\noindent where $n_{gas}(r)$ and $n_{CR}(r)$ are the densities of thermal gas and CR at radius $r$ from the cluster center, $v_{CR}\sim c$ is the velocity of the CR, $\sigma_{pp}$ is the proton-proton interaction cross section, and $E_{\gamma}$ is the energy of the produced photons. For the radial dependence of the CR density, we first assume that the CR follow the same radial distribution as the thermal gas (referred to as the ``isobaric'' case). As alternative possibilities, we also consider a profile where the CR-to-thermal energy ratio increases with radius as $\epsilon(r)\sim r^{0.5}$ (referred as ``flatter''), as expected from the acceleration of CR at cosmological shocks \citep[see e.g.,][]{Vazza12a, pinzke10,Donnert}, and a decreasing radial profile ($\epsilon(r)\sim r^{-0.5}$, referred as ``steeper'') to model the injection of CR by a central AGN \citep[e.g.,][]{WR, Mathews, Fujita}. Finally, upper limits on the CR energy density were derived for the three different cases and compared with the average thermal energy in our population, which was computed using the parameters given in \citet{chen07}. 

Among the 53 systems comprising our sample, 32 are classified as NCC, while 21 exhibit CC properties. As an estimate of the virial radius $R_{vir}$, we used the scaling relations of \citet{Arnaud}, making the approximation $R_{vir}\sim R_{200}$. We note that among our sample CC clusters typically exhibit a lower temperature ($T_{av,CC}\sim3.2$ keV), and thus a lower mass and thermal energy, than NCC systems ($T_{av,NCC}\sim5.5$ keV). 

\subsection{Results}
\label{sec:results}

Our results on the CR-to-thermal energy ratio are given in Table \ref{table:epsilon}. We report the results obtained for the entire sample, as well as for the CC and NCC cluster populations independently, for the three different CR radial distributions described above. As it can be seen in Table \ref{table:epsilon}, for a photon index of 2.0 our upper limits on the CR-to-thermal energy ratio for the whole sample are in the range 3-6\%. We stress that these values represent average values for the sample and cannot be used on a single-object basis. For comparison, using data obtained by the EGRET experiment on board the \emph{Compton Gamma-Ray Observatory}, \citet{reimer03} performed a similar study to the one presented in this paper, reaching upper limits on the CR-to-thermal energy ratio of the order of 10-20\% also for a photon index of 2.0. Thanks to the excellent sensitivity of Fermi-LAT, our upper limit lies a factor 3-4 below the upper limits obtained in this study. Recently, \citet{fermi10} performed a similar analysis on individual systems, always resulting in gamma-ray upper limits. This study allowed to constrain the CR energy density at the level of a few percent of the thermal energy density, in the best cases. Similar results were recently obtained on the Coma cluster through a combination of Fermi-LAT and VERITAS data \citep{COMA}, which lead to an upper limit on the order of 2\% on the CR-to-thermal pressure ratio. Compared to these studies, our analysis puts constraints on the average population which are slightly below the few best cases for individual systems. The stacking method therefore allows us to bring the constraints on the typical cluster population to the level of the few best individual cases. Recently, \citet{Dutson12} stacked a sample of 114 clusters including a central radio galaxy with Fermi-LAT, and did not find any evidence for a signal in the stacked population, in agreement with the results presented here.

In the fully relativistic case, the limits provided here can be readily transformed into the pressure ratio using the relation $\frac{U_{CR}}{U_{th}}=2\frac{P_{CR}}{P_{th}}$. This assumption is approximately valid, since we are considering only protons with kinetic energies $>1$ GeV. Thus in the isobaric case, for the average population we obtain an upper limit on the pressure ratio of $P_{CR}/P_{th}\lesssim2.2\%$. It has been claimed from numerical simulations \citep[e.g.,][]{Ando08} that the presence of a significant CR component in galaxy clusters could introduce a bias in cluster masses estimated assuming hydrostatic equilibrium. Our results indicate that the pressure contribution from CR is small, and thus that the bias in cluster masses induced by the presence of CR, if existing, would be negligible.

\begin{table}
\caption{\label{table:epsilon}Upper limits on the CR-to-thermal energy ratio of the parent proton population.}
\begin{center}
\begin{tabular}{cccccc}
\hline
$\Gamma_{ph}$ & $\Gamma_p$ & CR distribution & All & CC& NCC\\
\hline
\hline
2.0 & 2.05 & Isobaric       & $4.5$    &    $8.5$ & $3.4$ \\
 & & Steeper      & $3.1$     & $4.9$ & $2.6$  \\
 & & Flatter       & $6.2$    & $13.3$ & $4.5$ \\
2.4 & 2.45 & Isobaric       & $4.6$    &    $9.6$ & $3.3$ \\
 & & Steeper        & $3.2$     & $5.6$ & $2.4$\\
 & & Flatter        & $6.4$    & $15.1$ & $4.2$\\
2.8 & 2.85 & Isobaric       & $7.0$     &    $16.7$ & $4.3$\\
 & & Steeper         & $4.8$    & $9.7$ & $3.2$\\
 & & Flatter       & $9.6$     & $26.1$ & $5.6$\\
3.2 & 3.25 & Isobaric        & $10.1$    &    $30.7$ & $5.9$\\
 & & Steeper       & $7.4$      & $17.8$ & $4.4$\\
 & & Flatter        & $14.9$    & $47.8$ & $7.7$\\
\hline
\end{tabular}
\end{center}
\textbf{Column description:} 1: Photon index assumed for the analysis. 2: Corresponding effective proton index (see Fig. \ref{fig:spec_example}). 3: Assumed radial distribution of the CR in the cluster. In the ``isobaric" case, the CR distribution is assumed to follow the same radial dependence as the gas (Eq. \ref{gas_dist}). In the ``steeper" case the CR-to-thermal energy ratio decreases with radius as $\epsilon(r)\sim r^{-0.5}$. In the ``flatter" case it increases as $\epsilon(r)\sim r^{0.5}$. 4: UL on the CR-to-thermal energy ratio, in percent, computed using Eq.~\ref{av_flux} for the entire cluster population. 5: Same as 4 for the CC subsample. 6: Same as 4 for the NCC subsample.
\end{table}

\section{Discussion}
\label{sec:discussion}

\subsection{CR injection at merger shocks}
 
Numerical simulations modelling the formation of large-scale structures always report the presence of strong (i.e. with a Mach number $M \sim 10$) accretion shocks in the outer regions of galaxy clusters, and weaker ($2\leq M \leq 5$) and more energetic merger shocks induced by structure-formation processes in the innermost cluster regions \citep[e.g.,][]{Miniati00,Miniati01,Ryu,pfrommer08,Vazza12a}. These shocks are expected to inject a population of CR protons, which should accumulate in the cluster's volume since the formation epoch \citep{berezinsky97,voelk99}. The data explored here thus allows us to constrain the overall amount of CR protons injected at cosmological shocks. Recently, using pure hydrodynamical simulations \citep{Vazza12a} estimated that the CR-to-thermal pressure ratio should slightly increase with radius, from $\sim1\%$ in the core to $\sim10\%$ around $R_{200}$, without any important differences between relaxed and dynamically-active systems. Similar results were obtained by \citet{pinzke10}. Since the bulk of the CR are produced in merger shocks with Mach number $2-4$, we expect the photon index of the resulting proton population to be in the range 2.3 to 2.8. Comparing with our computation of the CR-to-thermal pressure ratio (see section \ref{sec:results} and Table \ref{table:epsilon}), for a flatter CR distribution we can see that our 95\% upper limits on the pressure ratio for the entire sample are in the range $3-5\%$ for $\Gamma_{p}$ in the range 2.3-2.8. These values are similar to the expectations of numerical simulations, and thus our observational results are starting to probe the particle acceleration models assumed in these simulations. For the NCC subpopulation, our upper limits are at the level of $2-3\%$ of the thermal pressure for $\Gamma_{p}$ in the range 2.3-2.8, which is in slight tension with the predictions. Therefore, although these predictions cannot be firmly ruled out yet, it is likely that the acceleration efficiency assumed in these simulations is overestimated.

Given that the simulations presented by \citet{Vazza12a} and \citet{pinzke10} are non-radiative and neglect the injection of CR by other processes (AGN, SNe), the predicted level of CR energy density should be treated as a lower bound to the expectations from numerical simulations, which reinforces our result. As discussed in \citet{Vazza12b}, a possible explanation for this result is the poorly-known acceleration efficiency of CR at weak shocks. Indeed, \citet{Vazza12a} used the efficiency of CR injection expected from DSA theory, following \citet{Kang}. Since the DSA theory is tailored to reproduce the acceleration efficiency of high-Mach number shocks in galactic supernova remnants, our results could indicate that CR acceleration at weak shocks is significantly less efficient than expected from DSA. Moreover, recent results on supernova remnants have shown that the accelerated CR spectra are steeper than expected from DSA \citep{Caprioli,Kang12}, even in the high-Mach number regime. Therefore, it appears likely that the acceleration efficiency used in existing numerical simulations is overestimated in the case of cosmological shocks.

Alternatively, cosmic rays could have been transported outwards to lower-density regions, which would make them unobservable to us \citep{ensslin11,keshet10}. Indeed, as we can see in Table \ref{table:epsilon}, our constraints are less tight when assuming a flat radial distribution for the CR, since the density of target protons drops sharply in the outer regions \citep{e12}. Therefore, an efficient transport of cosmic rays towards the outskirts could be responsible for making the CR distribution even flatter than considered here, resulting in a lower gamma-ray flux.

\subsection{AGN feedback in cool-core clusters}

One of the main discoveries of \emph{Chandra} has been the discovery of X-ray cavities at the center of relaxed galaxy clusters, which are thought to be inflated by powerful AGN outbursts \citep[see][for reviews]{mcnamara07,mcnamara12,fabian12}. Indeed, radio-loud AGN are found to be ubiquitous at the center of CC clusters \citep{Burns90,mittal09}, and in many cases CR electrons are filling the volume of X-ray cavities, providing strong evidence of the interplay between the central engine and the surrounding plasma. The energy injected by the central AGN can be large, preventing the gas from cooling below X-ray emitting temperatures and forming stars. While the general picture is clear, the details of the heat transfer mechanism are still not understood. One of the proposed mechanisms \citep[e.g.,][]{WR, Mathews, Fujita} considers the interaction of a population of relativistic CR with the gas as heat conveyor. In this framework, it has been shown that the observed X-ray properties can be recovered if the CR energy density is large enough. Our upper limit on the CR-to-thermal energy ratio in CC clusters thus allows us to put constraints on this model.

For a proton spectral index of 2.7, \citet{WR} found that they could explain the observed X-ray temperature profile (and the radio halo, when observed) in CC clusters if the CR radial distribution is steeper than the one of the gas and if the pressure ratio is on the order of $\sim0.4-1.2$, depending on the considered clusters. The typical expected gamma-ray fluxes exceed $10^{-10}$ ph cm$^{-2}$ s$^{-1}$ when converted in the [1-300] GeV band, which is several times larger than our upper limit. \citet{Fujita12} used a model where the CR are injected in the ICM during AGN intermittent explosions whose outgoing shock waves, followed by the formation of bubbles, heat the gas. This model was applied to the observations made on Perseus. Again, in this case a CR radial distribution more peaked than that of the gas and a pressure ratio in the order of 1-25\% (depending on the distance to the cluster center) were required to explain the observations. 

Comparing with our observational results, in the case of a CR profile decreasing with radius and for proton indices between 2.5 and 3.0, we obtain an upper limit on the CR-to-thermal pressure ratio of 3-5\% which is well below the values required to offset cooling. We note that because of the large target densities in the central regions, our analysis is very sensitive to the CR energy density in the inner regions, especially in the case of a steep CR radial profile. Therefore, we conclude that the energy density in the central regions of CC clusters is likely lower than what is needed in these models to offset radiative cooling, and heating through cavity expansion and/or shocks is preferred.

In this framework, we note that the upper limit obtained here also has implications on the typical gamma-ray luminosity of the central AGN itself. Indeed, in the case of Perseus \citep{ngc1275} a bright variable gamma-ray source was detected by Fermi-LAT, corresponding to the central AGN NGC 1275/3C 84 \citep[see also][]{Eckert09,Cola10,magic1275}. Our non-detection of the stacked CC cluster population with an upper limit $\sim3$ orders of magnitude below the gamma-ray flux of NGC 1275 thus shows that the central AGN of Perseus is significantly more active than the typical radio-loud AGN which are at work at the center of CC clusters. A similar conclusion was recently reached by \citet{Dutson12}.

\section{Summary}
\label{summary}
Making use of a maximum likelihood analysis of stacked Fermi-LAT count maps, we searched for gamma-ray emission from a sample of 53 galaxy clusters selected from the extended HIFLUGCS sample. Our results can be summarized as follows:
\begin{itemize}
\item
Assuming power-law spectra with a photon index of $\Gamma_{ph} = 2.0, 2.4, 2.8 \,\mathrm{and}\, 3.2$, we obtained a 95\% CL upper limits on the average photon flux per system (SUL/N) of few $10^{-11}\,\mathrm{ph}\,\mathrm{cm}^{-2}\,\mathrm{s}^{-1}$ in the [1-300] GeV energy band, for the total sample. This upper limit is an order of magnitude lower than the typical upper limits for individual clusters \citep{fermi10}.

\item
Performing separate analyses for the CC and NCC populations, we do not find evidence for differences in their gamma-ray emissivity. The 95\% CL flux upper limits (SUL/N) obtained for our CC and NCC sample are 5.7 and 2.7 $10^{-11}\,\mathrm{ph}\,\mathrm{cm}^{-2}\,\mathrm{s}^{-1}$, respectively, for a photon index of $\Gamma_{ph} = 2.4$.

\item
Assuming that the gamma-ray signal entirely comes from the decay of $\pi^0$ produced in $p-p$ collisions between CR protons and the ambient thermal gas, we derived upper limits of 4 to 10\% on the average CR energy density for a photon indices of $\Gamma_{ph}=2.0, 2.4, 2.8, \,\mathrm{and}\, 3.2$ (corresponding to a proton index of $\Gamma_{p}=2.05, 2.45, 2.85, \,\mathrm{and}\, 3.25$). Upper limits were also obtained for flatter and steeper radial distribution of the CR with respect to the thermal gas.

\item
Comparing our upper limits on the CR energy density with the expectations of numerical simulations modeling the injection of CR at cosmological shocks \citep{Vazza12a,pinzke10}, and taking into account that these simulations neglect the potential contribution of AGN and SNe to the global CR budget, our results are in tension with the CR level predicted by these simulations. Although we cannot completely rule out these models, this likely indicates that the injection of CR at low Mach number shocks is lower than expected from DSA. Alternatively, a very flat radial CR distribution (e.g., resulting from CR streaming), may reconcile our data with the expectations. 
\end{itemize}
\acknowledgements{We thank Franco Vazza and the anonymous referee for useful discussions and comments.}

\bibliographystyle{aa}
\bibliography{references.bib}

\normalsize

\onecolumn
\begin{center}
\begin{longtable}{c c c c c c}
\hline
Cluster & $l$ ($^\circ$) & $b$ ($^\circ$) & Redshift $z$ & $M_{500}$ (in 10$^{14}$ $M_{\odot}$) & Cool core\\
\hline
\endfirsthead

\multicolumn{6}{c}%
{\tablename\ \thetable\ -- \textit{Continued from previous page}} \\
\hline
Cluster & $l$ ($^\circ$) & $b$ ($^\circ$) & Redshift $z$ & $M_{500}$ (in 10$^{14}$ $M_{\odot}$) & Cool core\\
\hline
\endhead

\hline 
\multicolumn{6}{r}{\textit{Continued on next page}} \\
\endfoot

\endlastfoot
2A0335p096 & 176.25 & -35.07 & 0.0349 & 2.79 & yes\\
A0085 & 115.05 & -72.06 & 0.0556 & 8.08 & yes\\
A0119 & 125.70 & -64.10 & 0.0440 & 8.98 & --\\
A0133 & 149.76 & -84.23 & 0.0569 & 4.30 & yes\\
A0262 & 136.58 & -25.09 & 0.0161 & 0.94 & yes\\
A0399 & 164.36 & -39.47 & 0.0715 & 7.74 & --\\
A0401 & 164.18 & -38.87 & 0.0748 & 8.38 & --\\
A0478 & 182.41 & -28.30 & 0.0900 & 8.85 & yes\\
A0496 & 209.59 & -36.49 & 0.0328 & 4.81 & yes\\
A0548w & 230.49 & -25.26 & 0.0424 & 1.00 & --\\
A0576 & 161.42 & 26.24 & 0.0381 & 4.61 & --\\
A1060 & 269.63 & 26.51 & 0.0114 & 2.50 & --\\
A1367 & 235.31 & 73.01 & 0.0216 & 7.42 & --\\
A1413 & 226.19 & 76.78 & 0.1427 & 9.77 & --\\
A1644 & 304.90 & 45.50 & 0.0474 & 7.34 & yes\\
A1656 & 58.08 & 87.96 & 0.0232 & 9.95 & --\\
A1736 & 312.58 & 35.10 & 0.0461 & 2.17 & --\\
A1775 & 31.92 & 78.71 & 0.0757 & 4.19 & --\\
A1795 & 33.79 & 77.16 & 0.0616 & 9.87 & yes\\
A1800 & 40.47 & 77.07 & 0.0748 & 5.94 & --\\
A1914 & 67.20 & 67.46 & 0.1712 & 11.84 & --\\
A2142 & 44.23 & 48.69 & 0.0899 & 14.33 & --\\
A2151 & 31.58 & 44.52 & 0.0369 & 1.60 & yes\\
A2244 & 58.80 & 36.35 & 0.0970 & 5.48 & --\\
A2255 & 93.92 & 34.92 & 0.0800 & 7.86 & --\\
A2256 & 111.10 & 31.74 & 0.0601 & 12.12 & --\\
A2597 & 65.34 & -64.85 & 0.0852 & 3.71 & yes\\
A2634 & 103.45 & -33.06 & 0.0312 & 4.51 & --\\
A2657 & 96.65 & -50.30 & 0.0404 & 6.06 & --\\
A2877 & 293.13 & -70.88 & 0.0241 & 6.88 & --\\
A3112 & 252.95 & -56.09 & 0.0750 & 4.36 & yes\\
A3158 & 265.07 & -48.97 & 0.0590 & 5.75 & --\\
A3266 & 272.09 & -40.17 & 0.0594 & 19.24 & --\\
A3391 & 262.36 & -25.16 & 0.0531 & 6.04 & --\\
A3395 & 263.18 & -25.13 & 0.0498 & 9.48 & --\\
A3528n & 303.70 & 33.85 & 0.0540 & 4.49 & --\\
A3528s & 303.78 & 33.64 & 0.0551 & 2.76 & yes\\
A3530 & 304.00 & 32.51 & 0.0544 & 4.34 & --\\
A3532 & 304.44 & 32.48 & 0.0539 & 6.63 & --\\
A3558 & 311.98 & 30.74 & 0.0480 & 6.71 & --\\
A3560 & 312.73 & 29.00 & 0.0495 & 2.77 & --\\
A3562 & 313.31 & 30.35 & 0.0499 & 3.51 & --\\
A3571 & 316.32 & 28.55 & 0.0397 & 8.76 & --\\
A3581 & 323.13 & 32.85 & 0.0214 & 0.93 & yes\\
A3921 & 322.03 & -47.97 & 0.0936 & 6.59 & --\\
EXO0422m086 & 203.3 & -36.16 & 0.0390 & 2.72 & yes\\
Fornax & 236.72 & -53.63 & 0.0046 & 1.29 & yes\\
HydraA & 242.93 & 25.09 & 0.0538 & 4.07 & yes\\
MKW4 & 276.91 & 62.31 & 0.0200 & 0.69 & yes\\
NGC1550 & 190.98 & -31.85 & 0.0123 & 0.68 & yes\\ 
NGC499 & 130.50 & -28.94 & 0.0147 & 0.33 & yes\\
NGC5044 & 311.23 & 46.10 & 0.0090 & 0.49 & yes\\
NGC507 & 130.64 & -29.13 & 0.0165 & 0.46 & yes\\
\hline
\label{clustertable}\\
\caption{The sample of 53 clusters used for the stacking. The values for redshift $z$ and 
cluster mass $M_{500}$ are taken from \citep{chen07}. The classification of cool cores is done 
using \citep{cavagnolo09} and \citep{chen07}.}
\end{longtable}
\end{center}

\appendix

\section{Computation of the CR energy density}
\label{app:calculation}

\subsection{Proton-proton interaction model}
We consider a population of relativistic CR which interact with the ambient thermal gas. For each interaction, if the CR energy is larger than the threshold energy for pion production, its interaction with a thermal proton will produce pions and trigger a particle shower,
\begin{equation}
p+p\rightarrow n\pi^{0} + X,
\end{equation}
where  $n$ is the $\pi^0$ multiplicity and $X$ are the hadronic showers. The $\pi^{0}$ then decay into two photons, producing an observable signature in the gamma-ray range. We remark that in galaxy clusters, the gas density, magnetic fields, and radiation fields of the ambient medium are low such that all the produced pions will decay before interacting with the medium and the observed gamma-ray flux can be directly related to their parent protons.

In this work, we use the parametrization given by \citet{kelner_pp} to describe the produced photon spectrum. Since this parametrization is optimized for energies larger than 100 GeV, we apply the delta approximation at lower energy \citep[see][for details]{kelner_pp}.

The collision rate of protons with the gas can be described as
 \begin{equation}
dN/dt=n_{gas}n_{CR}v_{CR}(E_{p})\sigma_{pp}(E_{p}),
\end{equation}
where  $n_{CR}$ and $v_{CR}$ are the number density and velocity of the CRs, respectively, and $n_{gas}$ is the density of target protons. The interaction cross section, $\sigma_{pp}(E_{p})$, can be parametrized by the equation \citep{kelner_pp}

\begin{equation}\label{eq:sigma}
\sigma_{pp}(E_{p})=(34.3+1.88L+0.25L^2)\left[1-\left(\frac{E_{th}}{E_p}\right)\right] \mbox{mb},
\end{equation}
with $L=\ln(E_p/\mbox{1 TeV})$ and $E_{th}=m_p+2m_{\pi}+m^2_{\pi}/2m_p\sim1.22$ GeV, where $m_{\pi}$ and $m_p$ are the masses of the $\pi^0$ and the proton, respectively. It is logarithmically growing for energies larger than 100 GeV and approximatively constant in the range $E_p\in[5;100]$ GeV. In the energy range where the cross section is constant, the produced photon spectra follow the spectral shape of the parent proton population, whereas at lower energies, close to the threshold energy for pion production, the resulting spectra are harder.

\begin{figure}
  \begin{center}
    \includegraphics[width=0.5\columnwidth,angle=270]{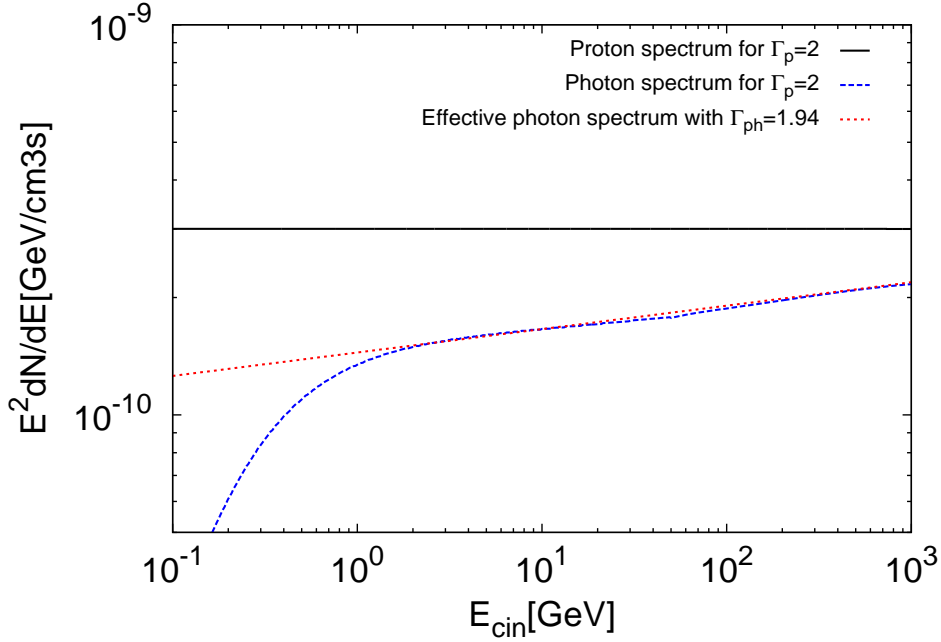}
  \end{center}
  \caption{The produced photon spectrum (blue) for a proton injected spectrum of $\Gamma_p=2$ and $E_{cut,p}=10^{15}$eV (black, scaled by an arbitrary factor). The red dotted line represents the best fit of the photon spectra with a simple power law in the energy range [1-300] GeV. The behavior at low energy follows the increase of the cross section at $E_p\sim1GeV$ (see Eq.~\ref{eq:sigma}). Note that the proton spectrum is expressed as function of the kinetic energy, $E_{cin}=E_{tot}-m_{p}c^2$, with $m_{p}\sim1 GeV/c^2$.}
  \label{fig:spec_example}
\end{figure}

The observational constraints from Fermi-LAT data were obtained by assuming that the spectral shape of the observed photons follows a single power law. However, below the threshold energy for pion production ($<$1 GeV), the predicted spectrum differs significantly from a power law. To alleviate this issue, we restrict to the photons with energies $>1$ GeV where the spectrum can be well-represented by a power law, and we fit the model spectrum with a simple power law over the energy range of interest. The best-fit photon index is the used as an effective photon index.  For instance, for a parent proton spectrum of $\Gamma$=2, we find an effective photon index of 1.94. This is illustrated in Fig. \ref{fig:spec_example}, where we represent in blue the photon spectrum produced by the interaction of the primary protons (in black) with the ambient gas. We can see that the effective photon index (red curve) is harder than the spectral index of the parent protons. Observationally, we obtained upper limits on the cluster emission for effective photon indices $\Gamma_{ph}=$2.0, 2.4, 2.8, and 3.2, which correspond to parent CR populations with $\Gamma_{p}$=2.05, 2.45, 2.85, and 3.25.

As for each interaction, 2 photons of energy $E_{\gamma}=\frac{\kappa}{2} E_{CR}$  are produced \citep[where $\kappa\sim0.17$ is the fraction of energy transferred from the proton to the pion $E_{\pi}=\kappa E_{CR}$,][]{kelner_pp}, the total gamma-ray luminosity is given by the product of the interaction rate with the photon energy,
 \begin{equation}\label{eq:lum}
L=2n_{gas}n_{CR}v_{CR}\sigma_{pp} E_{\gamma}.
\end{equation}

\subsection{Application to galaxy clusters}

We assume that the produced gamma-rays follow a power-law distribution with the same spectral index in all systems \citep[see][]{pinzke10}. In other terms, we assume that the CR are accelerated by the same mechanism, such that the CR content is expected to be of the same order of magnitude in all clusters. We assume that the CR distribution follows a power-law shape with an exponential cut-off at high energy \citep[choice motivated by the shock acceleration mechanism, e.g.][]{Bell},

\begin{equation}
\frac{dN_{CR}(r)}{dE_{CR}}=N_{0,CR}(r)\left(\frac{E_{CR}}{\mbox{1 TeV}}\right)^{-\Gamma_p}\exp\left(-\frac{E_{CR}}{E_{cut}}\right),\label{eq:crspectrum}
\end{equation}
where $\Gamma_p$ is the proton spectral index, $N_{0,CR}(r)$ is the normalization of the spectrum at distance $r$ from the cluster center, $E_{cut}$ is the cut-off energy. For proton indices larger than 2, the total CR energy density depends very weakly on the cut-off energy, and thus we fix $E_{cut}$ to $10^{19}$ eV. Since protons with kinetic energies $<1$ GeV are unobservable to us, we set a low-energy threshold of 1 GeV to the distribution.

For simplicity, we assume that the density of the thermal gas follows an isothermal beta model \citep{Cavaliere}, where the gas density can be parametrized by the function
 \begin{equation}\label{gas_dist}
  n_{gas}(r)= n_{0,gas}\left(1+\left(\frac{r}{r_c}\right)^2\right)^{-3\beta/2},
\end{equation}
 where $n_{0,gas}$ is the central density,  $r_{c}$ is the cluster core radius, and $\beta$ is the slope of the density profile in the outer regions. The thermal energy density at a radius $r$ is given by $u_{th}=\frac{3}{2}n_{gas}(r)kT$, where the temperature is assumed to be constant over the cluster volume. The total gamma-ray luminosity is then obtained by integrating Eq. \ref{eq:lum} over the entire cluster volume,
\begin{equation}
L=2\int^{R_{vir}}_{0}{n_{gas}(r)n_{CR}(r)v_{CR}\sigma_{pp} E_{\gamma}}\,\mbox{d}^3r.
\end{equation}
 
We express the total energy stored into CR as a fraction of the thermal energy, $U_{CR}=\epsilon\cdot U_{th}$. Given that CR accumulate in the cluster volume since the formation epoch, we expect little dependence of the CR energy density from one system to another. Thus, $\epsilon$ is assumed to be similar in all galaxy clusters, independently of their dynamical state. To determine $\epsilon$, we first assume that the CR are in equipartition with the gas (i.e. we set $\epsilon=1$) and determine the gamma-ray flux that should be observed under this assumption. The ratio of the observed UL to the flux expected when assuming equipartition then gives the value of $\epsilon$. 

Under the assumption of equipartition, we derive for each cluster the emitted photon spectrum per unit time and volume. The equipartition flux of cluster \textit{i} produced within these assumptions is then given by 
\begin{equation}
F_i=\frac{1}{4\pi d_{L,i}^2}\int_{V_i}\mbox{d}^3r \int_{1\mbox{\scriptsize{ GeV}}}^{300\mbox{\scriptsize{ GeV}}} E_{\gamma}\,\mbox{d}E_{\gamma} {\, \left[\frac{dN_{\gamma}}{dVdE_{\gamma}dt}\right]_{i}  }    .\label{eq:equipartitionflux}
\end{equation}
To compare with our upper limits on the stacked populations, we define the expected equipartition flux $F_{equip}$ as the mean of the equipartition fluxes of individual systems,
\begin{equation}\label{av_flux}
F_{equip}(\Gamma)=\frac{1}{N}\sum_{i=1}^{N}F_i,
\end{equation}
where $N$ is the total number of galaxy clusters in the sample. Then, an upper limit on the CR-to-thermal energy ratio can simply be obtained by taking the ratio of the observed flux to the equipartition flux,
\begin{equation}\label{factor}
\epsilon=\frac{{UL}_{Fermi}}{F_{equip}},
\end{equation}
where ${UL}_{Fermi}$ is the upper limit obtained in Sect.~\ref{results}.

\end{document}